# A Performance Comparison of Data Mining Algorithms Based Intrusion Detection System for Smart Grid

Zakaria El Mrabet [1], Hassan El Ghazi [2], and Naima Kaabouch [1]
[1] School of Electrical Engineering & Computer Science, University of North Dakota, USA
[2] National Institute of Posts and Telecommunication Rabat, Morocco

*Abstract*— Smart grid is an emerging and promising technology. It uses the power of information technologies to deliver intelligently the electrical power to customers, and it allows the integration of the green technology to meet the environmental requirements. Unfortunately, information technologies have its inherent vulnerabilities and weaknesses that expose the smart grid to a wide variety of security risks. The Intrusion detection system (IDS) plays an important role in securing smart grid networks and detecting malicious activity, yet it suffers from several limitations. Many research papers have been published to address these issues using several algorithms and techniques. Therefore, a detailed comparison between these algorithms is needed. This paper presents an overview of four data mining algorithms used by IDS in Smart Grid. An evaluation of performance of these algorithms is conducted based on several metrics including the probability of detection, probability of false alarm, probability of miss detection, efficiency, and processing time. Results show that Random Forest outperforms the other three algorithms in detecting attacks with higher probability of detection, lower probability of false alarm, lower probability of miss detection, and higher accuracy.

*Keywords*— Smart Grid, IDS, Data mining algorithms, probability of detection, probability of false alarm, Random Forest, Naïve Bayes

## I. INTRODUCTION

Smart grid is an emerging and promising technology. It uses the power of data communication to deliver intelligently the electrical power and meet the environmental requirements. Compared to the traditional power grid, smart grid provides new functionalities such as real-time control, two-way flow of information communication, operational efficiency, and grid resilience [1]. Because of the inherent vulnerabilities of communications systems, smart grid can be subject to a wide variety of security attacks. One of the systems for detecting malicious activity and securing the networks is the intrusion detection system (IDS).

IDS is based on three distinct approaches in detecting abnormalities: signature-based, specification-based, and anomaly-based. The first approach consists in detecting patterns of malicious activities using a database of well-known attack signatures. The second approach consists in identifying deviation from normal behavior profiles using logical specifications. The third approach consists in looking for deviations from normal behavior profiles using statistical measures [2]. According to Faisal et al. [3], anomaly-based technique is more applicable and suitable approach for smart grid than signature-based and specification-based techniques.

Despite its wide use, intrusion detection system (IDS) suffers from several performance limitations, including limited detection accuracy and high rate of false positive [4].

Thus, a number of researchers tried to improve its performance by proposing other techniques and algorithms. For example, Zhang et al. [5] proposed a distributed intrusion detection system for smart grid using artificial immune system and support vector machine algorithms. Ten et al. [6] proposed a substation detection algorithm that can be used to systematically extract malicious "footprints" of intrusion-based steps across substation networks. Berthier et al. [7] proposed a specification-based IDS for advanced metering infrastructure using sensors. Faisal et al. [3] proposed an anomaly based IDS for the advanced metering infrastructure (AMI) using several data stream mining techniques.

To the best of our knowledge, no previous work compared the performance of data mining algorithms used by anomaly-based IDS in smart grid networks. This paper fills this gap by evaluating the performance of four data mining algorithms which are: Naïve Bayes [8], Decision tree [8], Support Vector Machine [9], and Random Forest [10]. They have been used separately in several works with IDS, especially in the Smart Grid network, and they have produced satisfactory results. The benchmark NSL-KDD [11] dataset is selected for this evaluation. The evaluation is based on several metrics including the probability of detection, probability of false alarm, probability of miss detection, efficiency, and processing time.

The reminder of this paper is organized as follows. In section II, we describe the data mining algorithms used in this work. Section III presents the metrics used for evaluating the algorithms and discusses the results of the simulations. Some conclusions are drawn in the final section.

## II. DATA MINING ALGORITHMS

In this work, we selected one algorithm from each of the four categories: Bayes model [19], Decision tree model [20], function model [16], and ensemble model [14]. These algorithms are: Naïve Bayes (NB), Decision tree (DT), Support Vector Machine (SVM), and Random Forest (RF). These algorithms have been used separately in several works with IDS, especially in the Smart Grid network, and they have produced satisfactory results. A description of each of these algorithms is given below.

### A. Naive Bayes (NB)

Bayes network is a widely used graphical model to represent and handle uncertain information [15]. NB is a simple Bayes network with the "naive" assumption of independence between child nodes [8]. The steps for this algorithm are [9]:

1. Let *D* be a training set of tuple and their associated class labels. In our case, a tuple is a packet and it is

represented by an n-dimensional attribute vector $X = (x_1, x_2, x_3,..., x_n)$.

2. Suppose we have *m* classes, in our case *m=2* as we have two classes which are attack class and normal class. Given a tuple or a packet *X*, NB will predict that *X* belongs to the class with the highest posterior probability. In other words, *X* belongs to *Ci* if and only if:

$$P(C_i|X) > (P(C_j|X) \text{ for } 1 \leq j \leq m, j \neq i \quad (1)$$

So we maximize the $P(C_i|X)$ which is given by the Bayes Theorem:

$$P(C_i|X) = \frac{P(X|C_i) \cdot P(C_i)}{P(X)} \quad (2)$$

3. Because *P(X)* is constant for all classes, we remove it and we need to maximize only the $P(X|C_i).P(C_i)$. The prior probability $P(C_i)$ is given by:

$$P(C_i) = \frac{|C_{i,D}|}{|D|} \quad (3)$$

Where $|C_{i,D}|$ is the number of training tuple of class $C_i$ in *D*.

4. As NB assumes that there is no relationship among child nodes or attributes, the $P(X|C_i)$ is given by:

$$P(X|C_i) = \prod_{k=1}^{n} P(x_k|C_i) \quad (4)$$

Where $x_k$ is the value of attribute $A_k$ for tuple *X*. As the attributes $A_1, A_2....A_n$ in our data set are categorical, the $P(x_k|C_i)$ is the number of tuples of class $C_i$ in *D* having the value $x_k$ for $A_k$, divided by $|C_{i,D}|$.

*B. Decision tree (DT)*

The decision tree is a supervised learning algorithm used for addressing classification problem [21]. It consists of internal nodes, branches, and leaf nodes [14]. Each internal node represents a test on an attribute, the branch is the outcome of the test, and the leaf node is the class label to which a given object belongs. In order to classify a new object using the decision tree, two steps are required: creation and classification.

Several algorithms exist to create a decision tree including ID3, C4.5, and CART. In this work, we selected C4.5 (J48 in Weka) and it works as follows [14]:

Let *D* and *Attribute_list* are the inputs, where *D* is the training set and their associated class labels, in our case there are two class labels: attack and normal. *Attribute_list* is the list of attributes describing the tuples, in our case, these are the attributes describing the packets. Below is the basic algorithm for creating a decision tree using these inputs.

**Algorithm 1:** Create_Decision_Tree (D, attribute_list)
1. Create a node *N*
2. If packet in *D* are all of the same class *C*
   *N* = leaf node labeled with the class *C*
   return *N*
3. If the *attribute_list* is empty
   *N* = leaf node labeled with the majority class in *D*
   return *N*
4. Calculate the Gain ratio of each attribute in *attribute_list*
5. *N.test* = attribute with the best Gain ratio
   // the best means the highest value
6. For each outcome *j* in the splitting of *D*
   // partition the tuple and grow a sub trees for each
   //partition
7. if $D_j$ is empty
   *N* = leaf node labeled with the majority class in *D*
8. else
   *N* = Create_decision_tree ($D_j$, attribute_list)
   endFor
9. return *N*

C4.5 uses the Gain ratio (step 4) method for selecting the attribute that separates the given tuples according to class. It is defined as follows:

$$GainRatio(A) = \frac{Gain(A)}{SplitInfo(A)} \quad (5)$$

Where Gain (A) is given by:

$$Gain(A) = Info(D) - Info_A(D) \quad (6)$$

Where

$$Info(D) = -\sum_{i=1}^{m} p_i \log_2(p_i) \quad (7)$$

And

$$Info_A(D) = \sum_{j=1}^{v} \frac{|D_j|}{D} \times info(D_j) \quad (8)$$

Where, *Info(D)* denotes the average amount of information needed to identify the class label of a packet in *D*. *Pi* is the probability that a tuple in *D* belongs to the class *Ci*, and it is calculated by $|C_{i,D}| / |D|$. *m* is the number of classes. *InfoA(D)* is the information required to classify a tuple from *D* based on the partitioning by *A*.

The *SplitInfoA(D)* represents the information generated after dividing the training data set *D* into *v* partitions where each partition is the outcomes of a test on an attribute *A*. It is given by:

$$SplitInfo_A(D) = -\sum_{j=1}^{v} \frac{|D_j|}{D} \times \log_2(\frac{|D_j|}{D}) \quad (9)$$

After building the decision tree, the DT classifies a new tuple or packet by starting at the root node of the tree, evaluating the test, and taking the branch appropriate to the outcome. This process is continued until a leaf is encountered, at which time the new tuple is asserted to belong to the class named by the leaf [1].

## C. Support Vector machine (SVM)

SVM is a machine learning algorithm used for classification and regression [2]. It is based on a hyper line classifier, which separates and maximizes the margin between two classes [3].

Let the data set $D$ be given as $\{(x_1, y_1),(x_2, y_2),(x_3, y_3),...,(x_N, y_N)\}$ where $x_i$ is the set of training tuples with associated class label $y_i$. Each $y_i$ can take one of two values, either +1 or -1, corresponding to the class 'attack' and class 'normal' in our case.

SVM finds the best decision boundary to separate two classes by searching for the maximum margin hyper line [4]. A separating hyper line can be written as:

$$h(x) = w.x + b = \sum_{i=1}^{N} \alpha_i y_i (x_i, x) + b \quad (10)$$

Where $w$ is a weight vector $W$, $N$ is the number of attributes, and $b$ is a bias.

In case of non-linear data, we can first transform the data through non-linear mapping to another higher dimension space, and then use a linear model to separate the data. The mapping function is done by a kernel function $K$, and this time we will use the following formula [5]:

$$h(x) = w.x + b = \sum_{i=1}^{N} \alpha_i y_i K(x_i, x) + b \quad (11)$$

Where $K(x_i,x)$ is the kernel function.

After finding and drawing the separating hyper line, SVM classifies a new tuple or packet based on its position with respect to this hyper line. For example, if a tuple lies on or above the hyper line, it will belong to the class +1 (in our case the class 'attack'), and if it falls on or below the hyper line, it will belong to the class -1 (in our case the class 'normal').

## D. Random Forest (RF)

RF is a machine learning classifier which induces an ensemble of unpruned and randomized decision trees for classification and prediction [4]. Randomness is achieved either by selecting a bootstrap sample from the training data set or by selecting randomly a subset of attribute at each node of each decision tree [18]. The pseudo-code of the RF algorithm is described in Figure 1 [4, 18].

After building the learner, RF classifies a new object $x$ by aggregating the decisions (votes) over all the decision trees $T_i$ in the forest. The predicted class $T_i(x)$ of x determined by the tree $T_i$ is the class that occurs often in the random forest, and has the majority of the votes [18]. In our experiment, we used the default value of *Number_of_features* and *Number_of_trees* in Weka. *Number_of_feature* is given by log2M+1, where $M$ is the number of attributes in the data set, and *Number_of_feature* is equal to 10.

---

Building a bootstrapped sample $D_i$ from the data set $D$.
Creating a decision tree $T_i$ using the training data set $D_i$
For each node in the decision tree
    Define the number of attributes and select a
    random subset $(x_1, x_2,..., x_n)$, where $n$ is the
    *Number_of_features*.
For $i$ =1 to *Number_of_trees*
    Create a forest of trees $T_i$.

Fig. 1. Pseudo-code of the Random Forest classifier

## III. RESULTS AND DISCUSSION

For this work, we used an improved version of KDD cup 99 data set called NSL KDD. KDD cup 99 is a data set gathered at the MIT Lincoln Laboratory. It contains 5 209 460 network connections divided into two subsets: training subset with 4 898 431 connections and testing subset with 311 029 connections [17]. Each connection represents a packet that has 41 features, 34 are numerical and 7 are symbolic. Each packet is classified either as a normal traffic or as an attack. Attacks fall into the four categories: Remote to Local (R2L), Remote to Local (R2L), Denial of service (DOS), and Probing [7–9, 11]. Description of each attack's category along with some attack's examples is given in the Table I.

TABLE I. ATTACK'S CATEGORIES IN KDD CUP 99 AND NSL_KDD

| Attack's category | Description | Attack's Examples |
|---|---|---|
| Remote to Local (R2L) | Unauthorized access from a remote machine | Password guessing |
| User to Root (U2R) | Unauthorized access to local root privileges from a local unprivileged user | Rootkits, buffer overflow attack |
| DOS | Denial of service | Teardrop attack and Smurf attack |
| Probing | Surveillance and scanning | Scanning attack |

Although KDD Cup 99 is considered as a benchmark data set for assessing anomaly detection algorithms, it has a large number of redundant connections, especially a repeating record of DOS category. In addition, there is a bias distribution of the four categories which makes an accurate classification of the U2R and R2L difficult. These issues have been addressed in the NSL-KDD data set [6]. In this new version, many redundant and repeated records were removed, so the number of training and testing data were reduced 125 937 records and 22 544 records, respectively. In our experiment, we randomly selected 20% of this data set, which consists 25 191 connections. Table II presents some statistics about NSL KDD.

TABLE II. NUMBER OF NETWORK CONNECTIONS IN KDD CUP 99 AND NSL_KDD

| Attack's category | Number of network connections | | | |
|---|---|---|---|---|
| | KDD Cup 99 | | NSL KDD | |
| | Training | Testing | Training | Testing |
| DOS | 3 883 370 | 299 853 | 45 927 | 7 458 |
| R2L | 1 073 | 11 980 | 942 | 1 656 |
| U2R | 105 | 4 437 | 105 | 1 298 |
| Probing | 41 102 | 4 166 | 11 656 | 2 421 |
| Normal | 972 781 | 60 593 | 67 343 | 9 711 |
| Total | 4 898 431 | 311 029 | 125 973 | 22 544 |

To evaluate and compare the efficiencies of the four algorithms, metrics used are [7]:

- Probability of detection or true positive rate (*TPR*)
- Probability of false alarm or false positive rate (*FPR*)
- Probability of miss detection or false negative rate (*FNR*)

- Efficiency or accuracy
- Processing time.

The probability of detection or true positive rate (*TPR*) corresponds to the number of attack connections (packets) detected divided by the total number of attack packets. It is given by:

$$Probability\ of\ detection = \frac{TP}{TP+FN} * 100 \quad (12)$$

Where true positive (*TP*) corresponds to the number of attack connections (packets) correctly classified as attacks and false negative (*FN*) is the number of attack connections that are classified normal packets by the algorithm.

The probability of false alarm or false positive rate (*FPR*) is the percentage of connections that are incorrectly classified as attacks, while in fact they are normal. This probability is given by:

$$Probability\ of\ false\ alarm = \frac{FP}{FP+TN} * 100 \quad (13)$$

Where false positive (*FP*) is the number of normal connections classified as attacks and true negative (*TN*) is the number of normal connections classified normal connections by the algorithm.

The probability of miss detection or false negative rate (*FNR*) is the percentage of connections that are incorrectly classified as normal, while in fact they are attacks. It is given by:

$$Probability\ of\ miss\ Detection = \frac{FN}{TP+FN} * 100 \quad (14)$$

The efficiency or accuracy is the percentage of correct classified connection by the algorithm over the total number of connections. This efficiency is given by:

$$Efficiency = \frac{TP+TN}{TP+TN+FP+FN} * 100 \quad (15)$$

Where false positive (*FP*) is the number of normal connections classified as attacks. False negative (*FN*) refers to the number of attack connections wrongly classified as normal.

Finally, the detection or processing time is the time taken to process a certain number of connections and classify them as normal or attack packets.

Fig. 2 shows the probability of detection or true positive rate (*TPR*) with respect to the number of connections. It is apparent from this figure that Random Forest has the highest probability of detection, followed by Decision Tree, SVM, and then Naïve Bayes. For instance, with 2000 instances, the *TPR* of Random Forest is 98.7%, Decision Tree is 98%, SVM is 97.5%, and then Naïve Bayes is 90.2%. It is noticeable also that *TPR* is almost independent of the number of connections, except for Naïve Bayes which decreases.

Fig. 3 illustrates the probability of false alarm, false positive rate (*FPR*), as a function of the number of connections. As observed, the *FPR* decreases as the number of connections increases for all the algorithms except Naïve Bayes. It can also be seen that the Random Forest has the lowest *FPR* value while Naïve Bayes has the highest value. For example, with 5000 connections, the *FPR* of Random Forest is 0.5% while it is 9.3% for the Naïve Bayes.

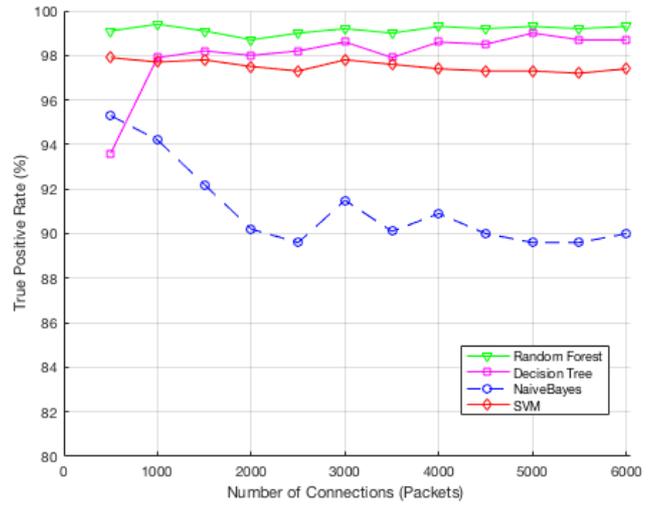

Fig. 2. True positive rate with respect to the number of connections

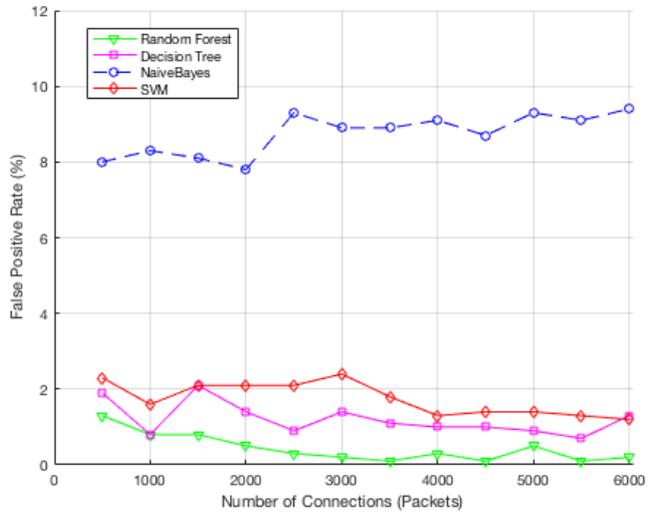

Fig. 3. False positive rate with respect to the number of connections

Fig. 4 presents the probability of miss detection or false negative rate (*FNR*) with respect to the number of connections. It is apparent from this figure that *FNR* remains almost the same for connection over 1000 for all algorithms except Naïve Bayes. The *FNR* of this algorithm increases sharply for connections under 2500 then slightly after. This figure also shows that Random Forest has the lowest *FNR* value, followed by Decision Tree, SVM, and then Naïve Bayes.

Fig. 5 illustrates the efficiency, the accuracy, of the four algorithms as a function of the number of connections. As one can see, Random Forest has the highest accuracy level, followed by Decision Tree, SVM, and then Naïve Bayes. For example, with 3500 connections, Random Forest reaches an accuracy of 99.48%, followed by Decision Tree with 98.41%, SVM with 97.89%, and then Naïve Bayes with 90.64%. Another noticeable thing is that the efficiency of Naïve Bayes decreases as the number of connections increases, whereas it is almost constant for the three other algorithms.

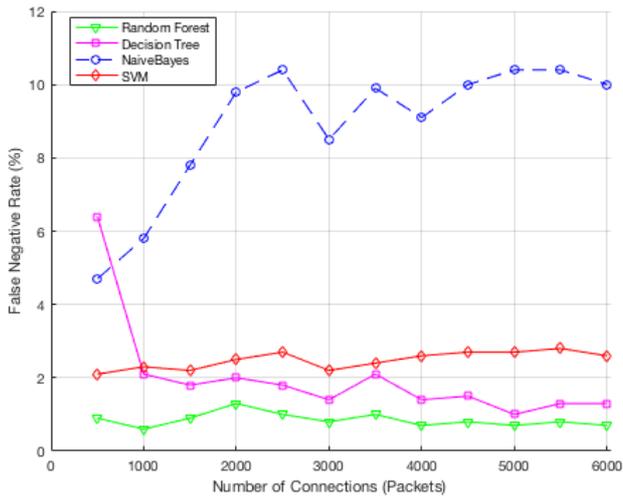

Fig, 4. False negative rate with respect to the number of connections

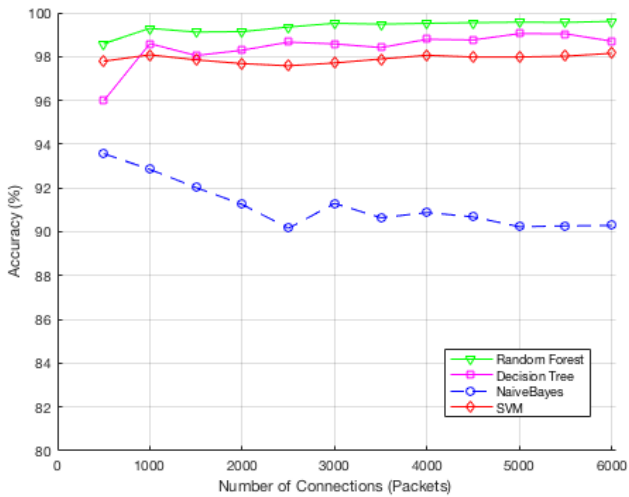

Fig. 5. Accuracy against the number of connections

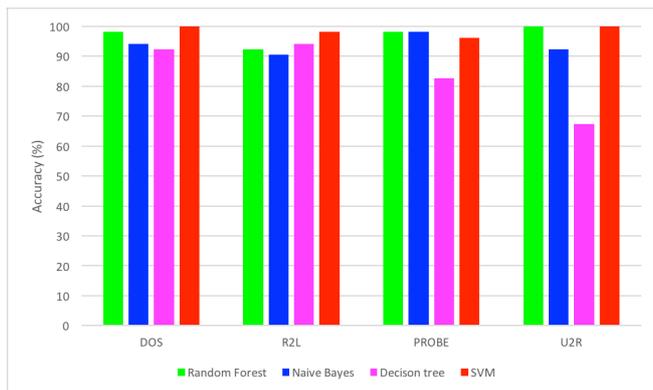

Fig. 6. Accuracy of four algorithms for each attack category

Fig. 6 shows the efficiency of the four algorithms for each attack category. Due to the bias distribution of the four attack categories in the NSL_KDD, we selected 208 instances for training and testing the algorithms, 52 instances of each category. As one can see, Random Forest detects efficiently (near to 100%) PROBE, U2R, and DOS attacks, while R2L attacks is better detected by SVM.

Fig. 7 shows the processing time as a function of the number of connections. As expected Random Forest and SVM require high processing time, followed by Decision Tree, and then Naïve Bayes. For instance, Random Forest takes 1.04s for processing 5500 connections, followed by SVM with 0.91s, Decision Tree with 0.21s, and then Naïve Bayes with 0.03s.

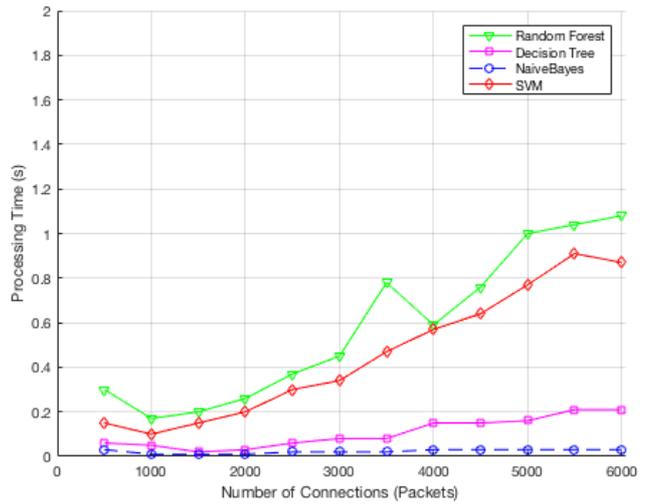

Fig. 7. Processing time against the number of connections

From these results, we can conclude that Random Forest is better than the other three algorithms in classifying attacks with higher probability of detection, lower probability of false alarm, lower probability of miss detection, and higher accuracy. It detects most of the attack including DOS, RPOBE, and U2R attacks. However, its main drawback is that it is time consuming. On the other hand, Naïve Bayes shows the lowest probability of detection, highest probability of false alarm, highest probability of miss detection, and lowest accuracy; but, it requires less processing time than the three other algorithms.

## IV. CONCLUSION

In this paper, we have evaluated the performance of four algorithms used by IDS in smart grid namely Random Forest, SVM, Decision Tree, and Naïve Bayes. For this assessment, we have selected several metrics, including probability of detection, probability of false alarm, probability of miss detection, efficiency, and processing time. The results show that some algorithms offer high accuracy with longer processing time, while others exhibit moderate accuracy but requires less processing time. As a future work, we aim to develop algorithms that satisfy the tradeoff between processing time and accuracy.